\documentclass[aps,superscriptaddress,amsmath,amssymb,floatfix,twocolumn,showpacs,amsfonts,longbibliography,reprint]{revtex4-2}
\usepackage{times}
\usepackage[varg]{txfonts}
\usepackage{textcomp}
\usepackage{graphicx}
\usepackage{subfigure}
\usepackage{color}
\usepackage[colorlinks=true,citecolor=blue,urlcolor=blue,linkcolor=blue,hyperindex]{hyperref}
\usepackage{braket}
\usepackage{float}
\usepackage{overpic}
\usepackage{bm}
\usepackage{array}
\usepackage{booktabs}
\usepackage{appendix}
\usepackage{tabularx}

\def\kk{\mathbf{k}}

\def\bs{\mathbf{S}}

\def\bq{\mathbf{Q}}
\def\bR{\mathbf{R}}

\begin{document}

\title{Raman scattering study of multimagnon (bi- and tri-magnon) excitations and rotonlike points in the distorted triangular lattice antiferromagnet}
\author{Junli Li}
\thanks{These two authors contributed equally.}
\affiliation{State Key Laboratory of Optoelectronic Materials and Technologies, Guangdong Provincial Key Laboratory of Magnetoelectric Physics and Devices, Center for Neutron Science and Technology, School of Physics, Sun Yat-Sen University, Guangzhou 510275, China}
\author{Shangjian Jin}
\thanks{These two authors contributed equally.}
% \affiliation{National University of Singapore, Singapore}
\affiliation{State Key Laboratory of Optoelectronic Materials and Technologies, Guangdong Provincial Key Laboratory of Magnetoelectric Physics and Devices, Center for Neutron Science and Technology, School of Physics, Sun Yat-Sen University, Guangzhou 510275, China}
\author{Trinanjan Datta}
\email[Corresponding author:]{tdatta@augusta.edu}
\affiliation{Department of Chemistry and Physics, Augusta University, 1120 15th Street, Augusta, Georgia 30912, USA}
\affiliation{Kavli Institute for Theoretical Physics, University of California, Santa Barbara, California 93106, USA}
\author{Dao-Xin Yao}
\email[Corresponding author:]{yaodaox@mail.sysu.edu.cn}
\affiliation{State Key Laboratory of Optoelectronic Materials and Technologies, Guangdong Provincial Key Laboratory of Magnetoelectric Physics and Devices, Center for Neutron Science and Technology, School of Physics, Sun Yat-Sen University, Guangzhou 510275, China}
\affiliation{International Quantum Academy, Shenzhen 518048, China}
%originating from the rotonlike M and M$^{\prime}$ points 
\begin{abstract}
We investigate the experimental signatures of Raman spectroscopy of bi- and tri-magnon excitations in the distorted triangular lattice antiferromagnets $\alpha$-LCr$_2$O$_4$  (L=Sr, Ca). Motivated by Raman scattering experiments, we utilize spin wave theory to analyze the nearly $120^\circ$ spin-3/2 spiral ordered antiferromagnetic ground state to compute the single-magnon density of states, single-magnon dispersion, and bimagnon and trimagnon Raman spectra (polarized and unpolarized). We perform calculations on the Heisenberg antiferromagnetic Hamiltonian that incorporates magnetic interactions (exchange, anisotropy, interlayer coupling) and lattice distortion within a four-sublattice unit cell. We investigate the Hamiltonian both for model parameter sets and experimentally proposed magnetic interactions for $\alpha$-LCr$_2$O$_4$  (L=Sr, Ca). It is found that Raman scattering is capable of capturing the effect of the rotonlike M and M$^{\prime}$ points on the bimagnon Raman spectrum. Our calculation confirms the connection between single-magnon rotonlike excitation energy and bimagnon Raman excitation spectrum observed experimentally. The roton energy minimum in momentum space is half of the energy of a bimagnon excitation signal. The experimental magnetic Raman scattering result displays two peaks which have a Raman shift of 15 meV and 40 meV, respectively. Theoretical modeling and analysis of the experimental spectrum of $\alpha$-SrCr$_2$O$_4$ within our distorted Heisenberg Hamiltonian lattice suggests that the low-energy peak at 15 meV is associated with the bimagnon excitation, whereas the high-energy peak around 40 meV is primarily a trimagnon excitation. Based on our fitting procedure we propose a new set of magnetic interaction parameters for $\alpha$-SrCr$_2$O$_4$. These parameters reproduce not only the experimental Raman spectrum, but also the inelastic neutron scattering response (including capturing high energy magnon branches). We also compute the unpolarized bimagnon and trimagnon Raman spectra for $\alpha$-CaCr$_2$O$_4$. In contrast to its Sr-cousin the Ca- based material has an enhanced bimagnon response, with the high energy peak still dominated by the trimagnon excitation. Furthermore, the polarization sensitivity of Raman spectrum can be utilized to distinguish the bi- and tri-magnon excitation channels.  
\end{abstract}
\maketitle
%%%%%%%%%%%%%%%%%%%%%%%%%%%%%%%%%%%%%%%%%%%%%%%%%%%%%%%
\section{Introduction}\label{sec:intro}
%%%%%%%%%%%%%%%%%%%%%%%%%%%%%%%%%%%%%%%%%%%%%%%%%%%%%%%

% \textbf{\datta{Change these wordings but first paragraph should be general motivation for triangular lattice and how quantum fluctuations are important}}

% \textbf{First para -} 
Experimental~\cite{JPSJ.56.4027,PhysRevB.81.104411,Ishii_2011,PhysRevLett.108.057205} and theoretical~\cite{PhysRevB.40.2727,PhysRevLett.68.1766,PhysRevB.50.10048,PhysRevB.79.144416,PhysRevB.79.184413,PhysRevB.81.020402,PhysRevLett.113.087204,PhysRevB.91.014426,PhysRevB.91.134423,PhysRevB.92.214409,PhysRevB.93.085111,PhysRevB.94.134416,PhysRevB.100.054410,PhysRevB.103.024417,Starykh_2015} studies of triangular lattice antiferromagnet (TLAF) suggest the stabilization of a long-range-ordered ground state that can be non-collinear and non-coplanar. The non-collinear ordering pattern is due to competing exchange interactions which in a perfect undistorted triangular lattice geometry manifests itself as a $120^{\circ}$ magnetic structure~\cite{PhysRevB.100.054410,PhysRevB.103.024417}. However, distortions of the underlying lattice network due to a heterogeneous magnetic unit cell modifies the strength of the exchange interactions. This in turn introduces anisotropic exchange interaction along various crystallographic directions of the lattice. The modified interactions result in a shift of the ordering wave vector, a behavior supported by theoretical investigation~\cite{PhysRevB.81.104411} and experimental observations~\cite{PhysRevB.81.104411,PhysRevB.84.054452,PhysRevLett.109.127203}. 
The TLAF has been investigated for its spin order and ground states using spin wave theory~\cite{PhysRevB.79.144416,PhysRevB.92.214409,PhysRevB.93.085111}. It has also been well studied theoretically and experimentally for its single-magnon excitation behavior using inelastic neutron scattering (INS) experiment~\cite{PhysRevB.88.094407,PhysRevB.35.4888}. Non-collinear magnetic ordering has the ability to harbor multimagnon excitations (bi- and trimagnon)~\cite{PhysRevB.92.035109,PhysRevB.100.054410,PhysRevB.103.024417}. The spectroscopic features of these multimagnon excitations have been investigated theoretically using RIXS at the $K$-edge~\cite{PhysRevB.92.035109}. Additionally, the multimagnon Raman bi- and tri- magnon excitation spectra have been investigated for the model undistorted TLAF compound using the torque equilibrium spin wave theory (TESWT)~\cite{PhysRevB.92.214409,PhysRevB.100.054410,PhysRevB.103.024417}. 
% \textbf{2nd para -} \datta{Summarize the RIXS and Raman findings of our calculations} 
In frustrated magnetic systems such as the TLAF, competition between exchange interaction, anisotropic XXZ interaction, or Dzyaloshinskii-Moriya (DM) interaction enhance the spin fluctuation and cause the failure of linear spin wave theory. The linear spin wave theory approach gives an incorrect ground state phase diagram while the $1/S$ corrected spin wave theory ($1/S$-SWT) gives an unphysical ordering wave vector. To remedy these problems, Du \emph{et al}.~\cite{PhysRevB.92.214409,PhysRevB.94.134416} established the TESWT which predicts a consistent result for the ground state phase diagram that agrees well with numerical calculations~\cite{PhysRevB.59.14367,Hauke_2011}. Furthermore, Ref.~\onlinecite{PhysRevB.100.054410} extended the TESWT approach to include the DM interaction that was well fitted to the INS experimental data of the spin spiral TLAF Cs$_2$CuCl$_4$~\cite{ColdeaPhysRevB.68.134424}. The authors computed the interacting bimagnon and non-interacting trimagnon RIXS spectrum and found that both spatial anisotropy and DM interaction modifies the RIXS spectrum at the two inequivalent rotonlike points, M$(0,2\pi/\sqrt{3})$ and M$^\prime(\pi, \pi/\sqrt{3})$. Specifically, the existence of DM interaction stabilizes the spiral state and increases the energy of roton minimum. 

Due to the issues related to experimental resolution, RIXS experimental studies on TLAF are non-existent. Compared to RIXS, Raman experiment which probe multimagnon excitations in TLAF are more prevalent~\cite{Wulferding_2012,PhysRevB.91.144411}. Bimagnon excitations have been observed with Raman spectroscopy in the distorted TLAFs -- $\alpha$-SrCr$_2$O$_4$~\cite{PhysRevB.91.144411} and $\alpha$-CaCr$_2$O$_4$~\cite{Wulferding_2012}. To study the magnon excitation spectrum in such spin spiral systems, Ref.~\onlinecite{PhysRevB.103.024417} presented a TESWT analysis that considered the anisotropic XXZ interaction and the DM interaction to compute the polarized Raman spectrum of Cs$_2$CuCl$_4$ and Ba$_3$CoSb$_2$O$_9$~\cite{PhysRevLett.110.267201}. It was reported that both the bimagnon and trimagnon excitations contribute to the Raman spectrum, which are influenced by spatial anisotropy, XXZ interaction and DM interaction. In contrast to DM interaction, XXZ interaction plays a weaker role in stabilizing the helical state.

When the spiral order is 120$^\circ$ (or approximately around this value), spin Casimir effect is negligible and TESWT reverts back to spin wave theory. Since the spiral order of $\alpha$-SrCr$_2$O$_4$ and $\alpha$-CaCr$_2$O$_4$ are close to 120$^\circ$, we apply linear spin wave theory in this paper to analyze this class of TLAF compounds. Experimental data on the distorted TLAF compounds $\alpha$-SrCr$_2$O$_4$ and $\alpha$-CaCr$_2$O$_4$ suggest spiral ordering temperatures below 43 K and 42.6 K, respectively~\cite{Dutton_2011,PhysRevB.91.144411,PhysRevLett.109.127203,Wulferding_2012}. For $\alpha$-SrCr$_2$O$_4$, Raman scattering experiment has reported the presence of bimagnon excitation.

In this article, we focus on the Raman spectrum of multimagnon excitations present in the the distorted TLAFs on which experimental data have been reported. We perform a spin wave analysis of the ordered spiral state to compute the effects of the rotonlike M and M$^{\prime}$ points on the Raman spectrum of the undistorted and distorted TLAFs. We study our model for a generic set of parameters to highlight the connection between the rotonlike points and their consequences on the bi- and tri-magnon Raman excitation spectrum. Based on experimental data we propose a new set of magnetic interaction parameters to compute the Raman spectrum of $\alpha$-SrCr$_2$O$_4$ and use existing ones to predict the Raman spectrum of $\alpha$-CaCr$_2$O$_4$. We show that polarized Raman spectroscopy has the ability to distinguish the bimagnon excitation channel from the trimagnon response. This difference is evident from a couple of perspectives. From an energetic point of view, the trimagnon signal always occurs at a higher energy compared to the bimagnon response. From a purely experimental scattering geometry set-up, we find that the HH signal is more sensitive to the trimagnon excitation, whereas the HV signal has a more pronounced bimagnon response. Our theory suggests that the Raman scattering experiment captures both the bimagnon and the trimganon excitation according to the spin dynamic features revealed by the INS experiment~\cite{PhysRevB.96.024416}.

This paper is organized as follows. In Sec.~\ref{sec:model}, we introduce the Heisenberg model of the distorted TLAF. In Sec.~\ref{sec:spin wave}, we compute the spin wave spectrum. In Sec.~\ref{sec:raman spectrum}, we use spin wave theory to compute and discuss the physical implications of the bimagnon and trimagnon Raman spectrum for our model Hamiltonian and for the real materials $\alpha$-LCr$_2$O$_4$  (L=Sr, Ca) (using experimental data). In Sec.~\ref{sec:conclusion}, we provide our conclusions. In the Appendices \ref{appendix a} and \ref{appendix b} we list the equations for the classical ground state energy analysis equation and the polarized Raman scattering operator matrix elements, respectively. 

\section{Model}\label{sec:model}
The spatially isotropic TLAF Ba$_3$CoSb$_2$O$_9$ exhibits a $120^\circ$ spin spiral order which is well described by a spin--1/2 XXZ model~\cite{PhysRevLett.110.267201,PhysRevLett.112.127203,PhysRevLett.112.259901,PhysRevB.91.024410,PhysRevLett.114.027201,PhysRevB.103.024417}. For the anisotropic TLAF, the ground states of Cs$_2$CuCl$_4$ and Cs$_2$CuBr$_4$ are long-range incommensurate spin spiral order in zero magnetic field~\cite{OnoPhysRevB.67.104431,PhysRevB.93.085111,ColdeaPhysRevB.68.134424}. The phases supported by these materials can be modeled using an antiferromagnet Heisenberg model with DM interaction~\cite{PhysRevB.100.054410,PhysRevB.103.024417,PhysRevB.75.174447}. The distorted TLAF $\alpha$-SrCr$_2$O$_4$ and $\alpha$-CaCr$_2$O$_4$ are reported to have approximate $120^\circ$ spin--3/2 spiral orders, with ordering wave vectors of $(\pi,\sim 4/3\times 2\pi,0)$~\cite{PhysRevB.84.054452,PhysRevB.96.024416,Dutton_2011}. In spite of the spiral ordering in these compounds, DM interaction is absent. 

%%%%%%%%%%%%%%%%%%%%%
\begin{figure}[b]
% \centering
\centerline{\includegraphics[width=8.0cm]{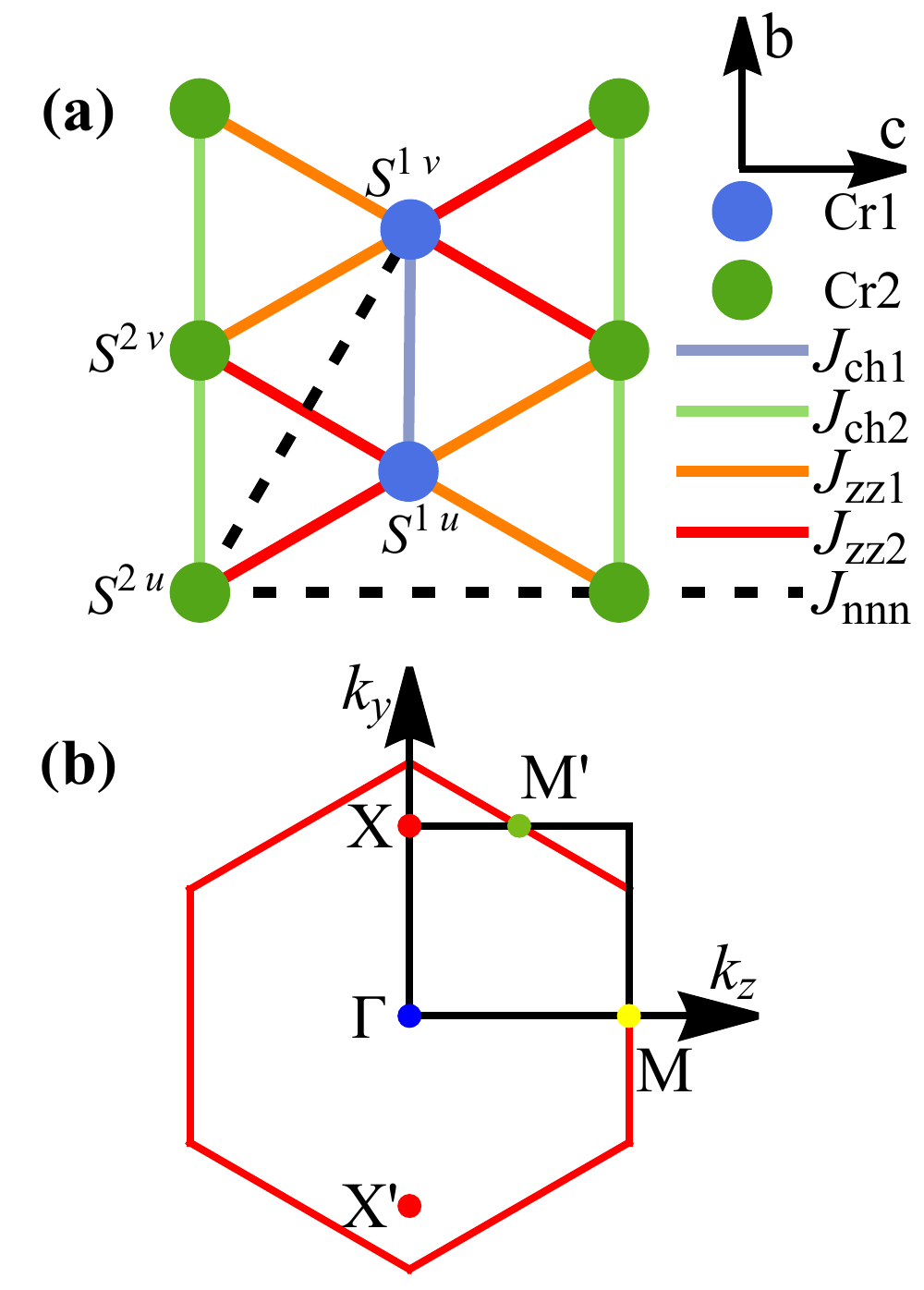}}
\caption{(a) Unit cell of $\alpha$-SrCr$_2$O$_4$~\cite{PhysRevB.96.024416} and $\alpha$-CaCr$_2$O$_4$~\cite{PhysRevB.84.054452}. The positions of the four Cr$^{3+}$ ions are $(0,0,0)$, $(0,0.5,0)$, $(0,0.25,0.5-0.01d)$, and $(0,0.75,0.5+0.01d)$ with site indices denoted by $i\in\{2u,2v,1u,1v\}$, where $d$ is the lattice distortion parameter. The two inequivalent Cr$^{3+}$ ions are shown as blue Cr1 and green Cr2 symbols. The spins are labeled as $\bs^{1u}$, $\bs^{1v}$, $\bs^{2u}$ and $\bs^{2v}$. (b) The black box is the first Brillouin Zone (BZ) of the unit cell with four spins. The red hexagon is the first BZ of the undistorted TLAF which contains only one spin in the unit cell. The high symmetry points $\Gamma$, M, M$^{\prime}$, $X$, and $X^{\prime}$ are shown in the BZ. The wave vectors are denoted by $k_y$ and $k_z$.}
\label{fig:fig1}
\end{figure}
%%%%%%%%%%%%%%%%%%%%%%%%%%%%%%%

The lattice structure of the distorted triangular lattice antiferromagnets $\alpha$-SrCr$_2$O$_4$ and $\alpha$-CaCr$_2$O$_4$ are shown in Fig.~\ref{fig:fig1}(a), which is an orthorhombic crystal structure belonging to the $Pmmn$ space group (no. 59). The lattice distortion parameter is $d=0.0025~(0.0057)$ for $\alpha$-SrCr$_2$O$_4$ ($\alpha$-CaCr$_2$O$_4$) as shown in Table.~\ref{table:1}. Such distortion results in a variety of exchange constants compared to the perfect TLAF, resulting in novel spin dynamic features in the presence of distorted TLAF. Each unit cell of $\alpha$-SrCr$_2$O$_4$ ($\alpha$-CaCr$_2$O$_4$) contains four atoms with two inequivalent magnetic ions Cr$^{3+}$. The distortion creates shifts on the sites, generating four different nearest-exchange constants. Former studies have shown that the Hamiltonian of the distorted TLAF contains nearest-neighbor (NN) and next-nearest-neighbor (NNN) interaction~\cite{PhysRevB.84.054452,PhysRevLett.109.127203,PhysRevB.96.024416}.

We consider a model which can be easily mapped to the undistorted triangular lattice by setting $d=0$. The Hamiltonian consists of the following interaction terms
\begin{equation}
\label{eq:ham}
H=H_{nn}+H_{nnn}+H_{int}+H_{an},
\end{equation}
where $H_{nn}$, $H_{nnn}$, $H_{int}$ and $H_{an}$ represent the nearest-neighbor exchange interaction, the next-nearest neighbor exchange interaction, the interlayer exchange interaction, and the single-ion anisotropy. The detailed expression for each interacting term is given by 
\begin{equation}
\label{eq:hnn}
\begin{split}
H_{nn}&=\sum\limits_{\alpha, \beta}J_{ch1}\bs^{1u}_{\alpha\beta}\cdot\left(\bs^{1v}_{\alpha\beta}+\bs^{1v}_{(\alpha-1)\beta}\right)+J_{ch2} \bs^{2u}_{\alpha\beta}\cdot\left(\bs^{2v}_{\alpha\beta}+\bs^{2v}_{(\alpha-1)\beta}\right)\\
&+J_{zz1} \bs^{1u}_{\alpha\beta}\cdot\left(\bs^{2u}_{\alpha(\beta-1)}+\bs^{2v}_{\alpha(\beta-1)}\right)
+J_{zz1} \bs^{1v}_{\alpha\beta}\cdot\left(\bs^{2u}_{(\alpha+1)\beta}+\bs^{2v}_{\alpha\beta}\right)\\
&+J_{zz2} \bs^{1u}_{\alpha\beta}\cdot\left(\bs^{2u}_{\alpha\beta}+\bs^{2v}_{\alpha\beta}\right)+J_{zz2} \bs^{1v}_{\alpha\beta}\cdot\left(\bs^{2u}_{(\alpha+1)(\beta-1)}+\bs^{2v}_{\alpha(\beta-1)}\right),
\end{split}
\end{equation}
\begin{equation}
\label{eq:hnnn}
\begin{split}
H_{nnn}&=\sum\limits_{\alpha,\beta}J_{nnn}\left[\bs^{2u}_{\alpha\beta}\cdot\left(\bs^{2u}_{\alpha(\beta+1)}+\bs^{1v}_{\alpha\beta}\right.\right.\\
&\left.+\bs^{1v}_{\alpha(\beta+1)}+\bs^{1u}_{(\alpha-1)(\beta+1)}+\bs^{1u}_{(\alpha-1)\beta}\right)\\
&+\bs^{2v}_{\alpha\beta}\cdot\left(\bs^{2v}_{\alpha(\beta+1)}+\bs^{1u}_{(\alpha+1)\beta}\right.\\
&\left.+\bs^{1u}_{(\alpha+1)(\beta+1)}+\bs^{1v}_{(\alpha-1)\beta}+\bs^{1v}_{(\alpha-1)(\beta+1)}\right)\\
&\left.+\bs^{1u}_{\alpha\beta}\cdot \bs^{1u}_{\alpha(\beta+1)}+\bs^{1v}_{\alpha\beta}\cdot \bs^{1v}_{\alpha(\beta+1)}\right],
\end{split}
\end{equation}
\begin{equation}
\label{eq:hint}
\begin{split}
H_{int}=J_{int}\sum\limits_{i,j}\bs_{\alpha\beta}\cdot \bs_{\alpha\beta+a},
\end{split}
\end{equation}
\begin{equation}
\label{eq:haniso}
\begin{split}
H_{an}=A\sum_{\alpha\beta}{\left( S_{\alpha\beta}^b \right)^2}.
\end{split}
\end{equation}
The four different spin labels $S^{2u}$, $S^{2v}$, $S^{1u}$ and $S^{1v}$ represent the four different sublattices in the unit cell of the distorted TLAF. The parameters $J_{ch1}$, $J_{ch2}$, $J_{zz1}$, $J_{zz2}$ corresponding to the exchange constants for $nn$ interaction are in the $(b,c)$ plane. The three other parameters $J_{nnn}$, $J_{int}$, $D$ are for the second $nn$ interaction, interlayer exchange interaction, and easy plane anisotropy, respectively. The $\alpha$ and $\beta$ in $\bs_{\alpha\beta}$ are sites of the unit cell along the $b$ direction and the $c$ direction, respectively. See Table~\ref{table:1} for parameter values. 

\section{Spin wave spectrum}\label{sec:spin wave}   
Inelastic neutron scattering experiments have been performed on $\alpha$-LCr$_2$O$_4$  (L=Sr, Ca) to study their spin wave dynamics~\cite{PhysRevB.96.024416,PhysRevLett.109.127203}. The neutron scattering study provides us with the exchange constants and anisotropy parameters of these materials ~\cite{PhysRevB.96.024416,PhysRevLett.109.127203}. We compute the magnon dispersion for the spin-spiral ground state in the distorted TLAF which exhibits the ordering wave vector $\bq$ which rotates in the $(a,c)$ plane to generate a helical spin order in its ground state configuration. The Hamiltonian in the local rotating basis is given by 
\begin{equation}
\begin{split}
H&=\sum\limits_{i, j}J_{ij}\left[S^y_iS^y_j+\cos\left(\bq\cdot\bR_{ij}\right)\left(S^x_iS^x_j+S^z_iS^z_j\right)\right.\\
&\left.+\sin\left(\bq\cdot\bR_{ij}\right)\left(S^z_iS^x_j-S^x_iS^z_j\right)\right]+A\sum\limits_i (S_i^y)^2,
\end{split}
\end{equation}
where $\bR_{ij}=\bR_i-\bR_j$ and $J_{ij}$ is the exchange interaction between sites $i$ and $j$. The first summation term contains the nearest neighbor, the next-nearest neighbor, and the interlayer exchange interactions, while the second summation contains the single-ion anisotropy. The spin components in the lab frame $(a,b,c)$ were transformed into the rotating local frame basis $(x,y,z)$ using the transformation~\cite{PhysRevB.79.144416}
\begin{equation}
\bs_i=\left(\begin{array}{ccc}
\cos(\bq\cdot\bR_i )  & 0 & \sin(\bq\cdot\bR_i)\\
0 & 1 & 0\\
-\sin(\bq\cdot\bR_i) & 0 & \cos(\bq\cdot\bR_i )
\end{array}
\right)
\left(\begin{array}{c}S_i^x\\S_i^y\\S_i^z\end{array}\right).
\end{equation}
\begin{table*} 
\begin{center}   
\caption{The parameter set choices for different distortion values utilized to compute the Raman spectra. $\mathcal{P}_4$ and $\mathcal{P}_5$ are the experimental (exp) parameter sets for the distorted TLAFs $\alpha$-LCr$_2$O$_4$  (L=Sr, Ca)~\cite{PhysRevB.96.024416,PhysRevLett.109.127203}. The penultimate line in the table reports the fit parameters generated from our model based on the experimental data of Ref.~\onlinecite{PhysRevB.91.144411}. The ordering wave vector $\bq=(Q_a,Q_b,0)$ with $Q_a =\pi$ and $Q_b$ varies based on different parameter sets. $\mathcal{P}_1$, $\mathcal{P}_2$ and $\mathcal{P}_3$ are the model parameter sets, representing the $120^\circ$ helical state. Below the N\'eel ordering temperature $T_N=43$ K $\alpha$-SrCr$_2$O$_4$ forms an incommensurate helical order~\cite{PhysRevB.91.144411,ChaponPhysRevB.83.024409,PhysRevB.84.054452,Dutton_2011}.For $\alpha$-CaCr$_2$O$_4$ the helical state forms below 42.6 K~\cite{PhysRevLett.109.127203}.\\}
\label{table:1}
\begin{tabular}{p{0.3\linewidth}m{0.063\linewidth}m{0.063\linewidth}m{0.063\linewidth}m{0.063\linewidth}m{0.063\linewidth}m{0.063\linewidth}m{0.063\linewidth}m{0.063\linewidth}m{0.063\linewidth}}
%\begin{tabular}{c|ccccccccc}
\toprule[0.3mm]
\toprule[0.3mm]
\textbf{Parameter set} & \textbf{$J_{ch1}$} & \textbf{$J_{ch2}$} & \textbf{$J_{zz1}$} & \textbf{$J_{zz2}$} & \textbf{$J_{nnn}$} & \textbf{$J_{int}$} & \textbf{$A$} & \textbf{$d$} & \textbf{$Q_b/2\pi$}\\ 
\midrule[0.1mm]
$\mathcal{P}_1$ (model) & 1.0 & 1.0 & 1.0 & 1.0  & 0.01 & 0.01  & 0.01 & 0      & $4/3$ \\ 
$\mathcal{P}_2$ (model) & 1.0 & 1.0 & 0.9 & 1.1  & 0.01 & 0.01  & 0.01 & 0.1  & $4/3$ \\  
$\mathcal{P}_3$ (model) & 1.0 & 1.0 & 0.8 & 1.2  & 0.01 & 0.01  & 0.01 & 0.2  & $4/3$ \\
$\mathcal{P}_4$ (exp, $\alpha$-SrCr$_2$O$_4$)~\cite{PhysRevB.96.024416} & 5.2 & 4.9 & 3.8 & 6.0  & 0.35 & 0.02  & 0.01 & 0.25 & 1.3217      \\
$\mathcal{P}_5$ (exp, $\alpha$-CaCr$_2$O$_4$)~\cite{PhysRevLett.109.127203} & 9.1 & 8.6 & 5.8 & 11.8 & 0.57 & 0.027 & 0    & 0.57 & 1.3317      \\
$\mathcal{P}_6$ (proposed fit, $\alpha$-SrCr$_2$O$_4$) & 6.5 & 7.5 & 4.6 & 7.9 & 0.45 & 0.02 & 0.06 & 0.25 & 1.3217 \\
Valentine et al. (exp, $\alpha$-SrCr$_2$O$_4$)~\cite{PhysRevB.91.144411} & 7.15 & 4.22 & 3.02 & 5.70 & -- & -- & -- & 0.25 & 1.322 \\
\bottomrule[0.2mm]
\bottomrule[0.2mm] 
\end{tabular}   
\end{center}   
\end{table*}
Next, we perform the Holstein-Primakoff (HP) transformation to recast the spin label $S^{x,y,z}_i$ to the quasiparticle label given by 
\begin{equation}\label{eq:HP}
 S_i^z=S-a_i^\dag a_i,\ S_i^-=a^\dag\sqrt{2S-a_i^\dag a_i},\ S_i^+=(S_i^-)^\dag,
\end{equation}
where $a_i^\dag$ ($a_i$) is the magnon creation (annihilation) operator for a given site $i$. The quadratic spin wave Hamiltonian $H_2$ can then be expressed as
\begin{equation}
\begin{split}
\label{eq:h2}
H_2&=\sum\limits_{i,j}J_{ij}S\left[\frac{1}{2}\left(-a_ia_j+a_ia^{\dagger}_j+h.c.\right)\right. \\
&+\frac{1}{2}\cos\left(\bq\cdot \bR_{ij}\right)\left(a_ia_j+a_ia^{\dagger}_j+h.c.\right) \\
&\left.-\cos\left(\bq\cdot \bR_{ij}\right)\left(a^{\dagger}_ia_i+a^{\dagger}_ja_j\right)\right] \\
&+\frac{AS}{2}\sum\limits_{i}(-a_ia_i+a_ia^{\dagger}_i+h.c.).
\end{split}
\end{equation}
After Fourier transformation the two interaction terms in $H_2$ are given by 
\begin{equation}
\begin{split}
\label{eq:h2f}
H_2&=\sum\limits_{\kk}\left\{\sum\limits_{i,j}J_{ij}S\left[\frac{1}{2}(-e^{i\kk\cdot \bR_{ij}}a_{\kk i}a_{-\kk j}+e^{i\kk\cdot \bR_{ij}}a_{\kk i}a^{\dagger}_{\kk j}+h.c.)\right.\right. \\
&+\frac{1}{2}\cos\left(\bq\cdot \bR_{ij}\right)(e^{i\kk\cdot \bR_{ij}}a_{\kk i}a_{-\kk j}+e^{i\kk\cdot \bR_{ij}}a_{\kk i}a^{\dagger}_{\kk j}+h.c.) \\
&\left.-\cos\left(\bq\cdot \bR_{ij}\right)(a^{\dagger}_{\kk i}a_{\kk i}+a^{\dagger}_{\kk j}a_{\kk j})\right] \\
&\left.+\frac{AS}{2}\sum\limits_{i}(-a_{\kk i}a_{-\kk i}+a_{\kk i}a^{\dagger}_{\kk i}+h.c.)\right\}.
\end{split}
\end{equation}
The exchange interaction $J_{ij}$ includes $J_{ch1}$, $J_{ch2}$, $J_{zz1}$, $J_{zz2}$, $J_{nnn}$, and $J_{int}$. Utilizing a numerical Bogoliubov transformation~\cite{van1980note}
\begin{equation}
\begin{pmatrix}
a_{\kk i} \\
a^\dag_{-\kk i}
\end{pmatrix}=
\begin{pmatrix}
u_{\kk ip} & v_{\kk ip} \\
v_{\kk ip} & u_{\kk ip}
\end{pmatrix}
\begin{pmatrix}
b_{\kk p} \\
b^\dag_{-\kk p}
\end{pmatrix}=
T_\kk
\begin{pmatrix}
b_{\kk p} \\
b^\dag_{-\kk p}
\end{pmatrix},
\end{equation}
we diagonalize the Hamiltonian Eq.~\eqref{eq:h2f} to obtain the spin wave dispersion
\begin{equation}
T^\dag_\kk H_2T_\kk=\text{diag}\left(\omega_{\kk i},-\omega_{\kk i}\right),
\end{equation}
diag represents the diagonal matrix of energy eigenvalues. As the ground states of $\alpha$-LCr$_2$O$_4$  (L=Sr, Ca) are both close to a $120^\circ$ order, we study three sets of model parameters ($\mathcal{P}_1$, $\mathcal{P}_2$ and $\mathcal{P}_3$) with different distortions $d$ and ordering wave vector $\bq$ (see Table~\ref{table:1} for parameter choices). Assuming that exchange interaction is inversely proportional to the bond length, we model the anisotropic interaction as $J_{zz2}$ ($J_{zz1}$) $\approx J_{ch1}(1 \pm d)$ according to the parameters of $\alpha$-LCr$_2$O$_4$ (L=Sr, Ca).

\begin{figure*}
% \centering
\centerline{\includegraphics[width=17.4cm]{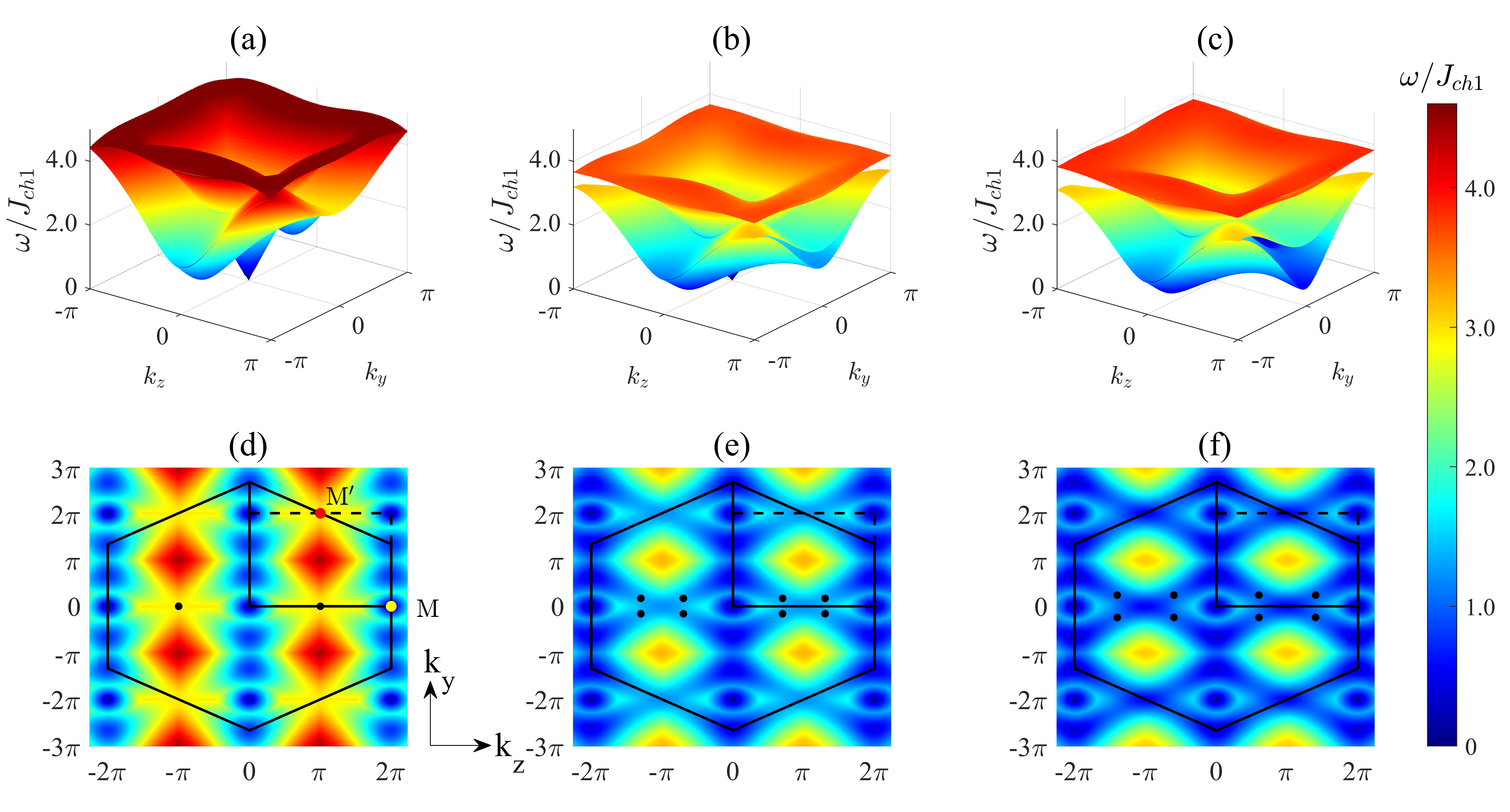}}
\caption{Spin wave dispersion of the triangular lattice antiferromagnet. (a)~-~(c) are spin wave dispersions for parameter sets $\mathcal{P}_1$, $\mathcal{P}_4$, and $\mathcal{P}_5$, respectively. There are four branches of energy for each parameter set as the unit cell contains four sublattices. (d)~-~(f) are one of the four energy branches with the lowest energy which correspond to (a)~-~(c). M$(0,0,2\pi)$ and M$^\prime(0,\pi,2\pi)$ are the rotonlike points of the undistorted triangular antiferromagnet for $\mathcal{P}_1$. The roton minimum energy can be observed in rotonlike points indicated by the black dots. We observe the roton energy minimum at different spin wave dispersion branches for various parameter sets, see Table~\ref{table:1}. Distortion results in M$^{\prime}$ becoming the only rotonlike point which splits into four different points in the spin wave dispersion. Enhanced distortion spreads the split rotonlike points (indicated by the eight black dots in panel (e)) into four different directions in momentum space (see panel (f)).}
\label{fig:fig2}
\end{figure*}

The single-magnon density of states (DOS) for $\mathcal{P}_1$, $\mathcal{P}_4$ and $\mathcal{P}_5$ in Table.~\ref{table:1} are calculated. In Figs.~\ref{fig:fig2}(a)~-~(c) we show the spin wave spectra for parameters sets $\mathcal{P}_1$ for the undistorted TLAF, $\mathcal{P}_4$ for $\alpha$-SrCr$_2$O$_4$, and $\mathcal{P}_5$ for $\alpha$-CaCr$_2$O$_4$. Inspecting the figures we can conclude that the local minimum points in the magnon dispersion are shifted by the presence of lattice distortion. It is reported that the local minima of the undistorted TLAF appear at the M and M$^\prime$ points, which are described as rotonlike points in literatures~\cite{PhysRevB.74.180403,PhysRevB.74.224420,PhysRevLett.96.057201}. However, in the presence of distortion, the local minima points are shifted, see the black dots in Figs.~\ref{fig:fig2}(d)-(f). Raman detection of rotonlike modes have been performed by Dirk \emph{et al}.~\cite{Wulferding_2012} in $\alpha$-CaCr$_2$O$_4$. The Raman spectrum shows a maximum around twice the roton energy. This is because the rotonlike points contribute to van-Hove singularities, causing a maxima in the DOS for one magnon. Since the bimagnon Raman signal comes from bimagnon excitation which creates two-magnons with inverse momentum, the maxima in Raman spectrum should reflect twice the energy of rotonlike points. However, the trimagnon Raman signal is from three magnon excitation with zero total momentum, $\kk_1+\kk_2+\kk_3$~=~0. The choice of wave vector for the trimagnon is not unique. Thus, the trimagnon Raman spectrum should be even more broader. 

The rotonlike points do not necessarily occur in the lowest energy branch. In Figs.~\ref{fig:fig2}(a) and (d), the M point of the undistorted TLAF shows the roton minimum in the second energy band. In contrast, the roton minimum for the 
 M$^\prime$ point always occurs in the lowest energy band. In Fig.~\ref{fig:fig2}(d), the M$^\prime$ point shifts to the black points in the first BZ of the distorted TLAF model due to periodic condition. Each rotonlike point splits into four rotonlike points as shown in Fig.~\ref{fig:fig2}(e) with enhanced distortion. Fig.~\ref{fig:fig2}(f) indicates that the rotonlike points for the distorted TLAF tend to spread away when distortion $d$ is increased. Finally, note that we have not considered magnon interactions. The Heisenberg model was treated at the quadratic linear spin wave level since the system is a spin-3/2 material. Thus, we expect quantum fluctuations to be suppressed. Additionally, the presence of multiple sublattices in the unit cell increases the algebraic complexity to pursue a fully interacting magnon calculation without contributing any additional understanding of the underlying spin dynamics behavior of the distorted TLAF.

\section{Raman spectrum}\label{sec:raman spectrum}
In this section we investigate the Raman spectrum of TLAF. We construct the Raman scattering operator for the bimagnon and trimagnon excitations. First, we perform calculations of the polarized Raman spectrum of $\alpha$-SrCr$_2$O$_4$ and $\alpha$-CaCr$_2$O$_4$. Second, we give the physical implications of the bimagnon Raman signal and the rotonlike points in the single-magnon spin wave dispersion. We also discuss the trimagnon response. Finally, we fit the experimental data of $\alpha$-SrCr$_2$O$_4$ with our unpolarized Raman spectrum and compute the unpolarized Raman spectrum of $\alpha$-CaCr$_2$O$_4$.
\subsection{Raman scattering operator}\label{sec:raman scattering operator}
\begin{figure*}
\centerline{\includegraphics[width=16.8cm]{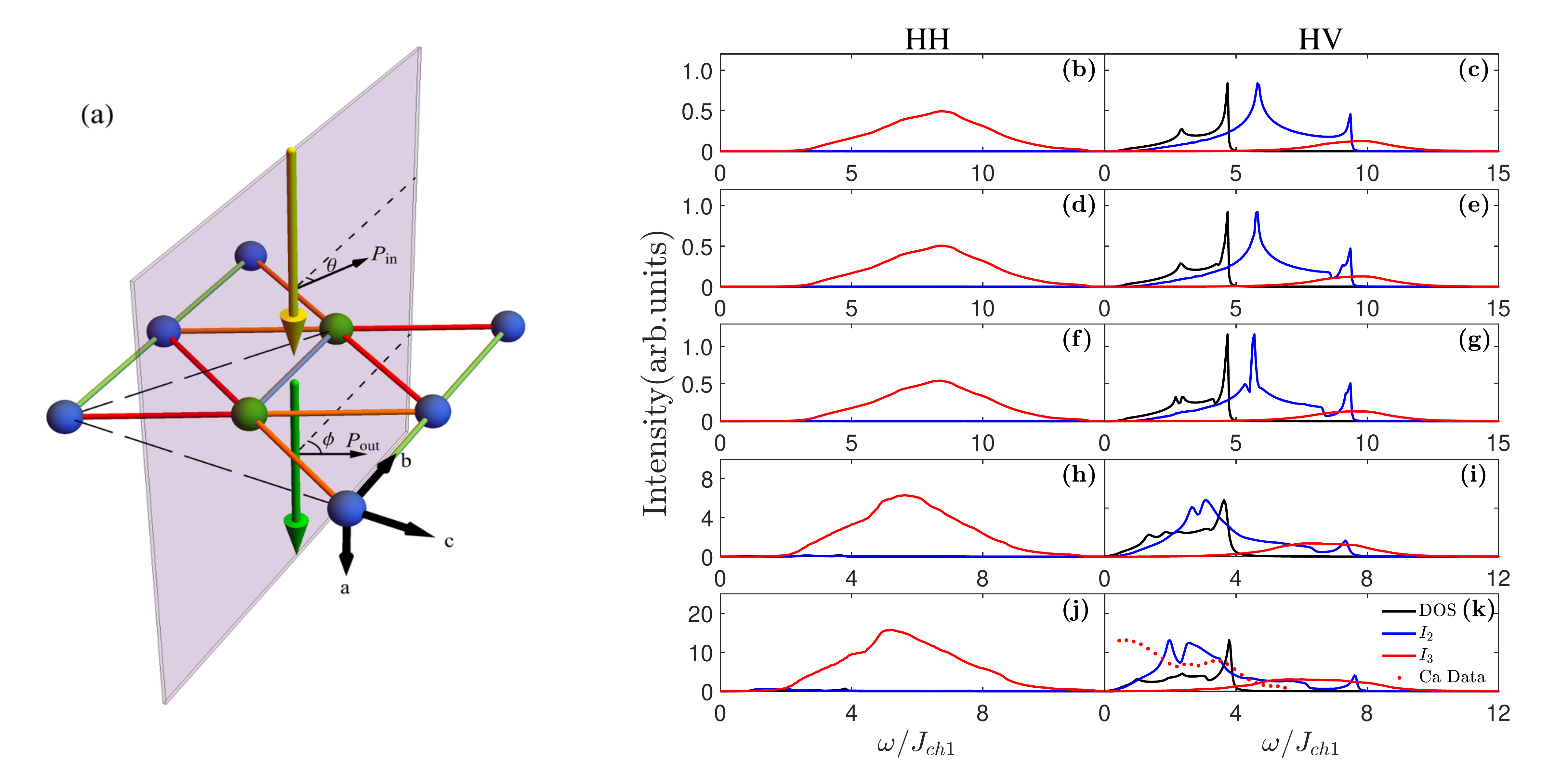}}
\caption{(a) Raman scattering geometry. Incident light is always assumed to be along the $a$ axis. The angle $\theta$ and $\phi$ are defined as displayed in the figure. (b)-(k) present the polarized bimagnon $I_2(\omega,\theta,\phi)$ and the trimagnon $I_3(\omega,\theta,\phi)$ Raman spectra. Each row refers to the parameter choice $\mathcal{P}_{i}$ with $i=1,\dots, 5$, respectively (see Table~\ref{table:1}). The first column represents HH polarization with $(\theta,\phi)=(0,0)$. The second column is HV polarization with $(\theta,\phi)=(0,\frac{\pi}{2})$. All the calculated results are computed by applying $p_1=p_2=p_3=p_4=1$. The black solid line is single-magnon density of states. Red dots are extracted using WebPlotDigitizer~\cite{Rohatgi2022} from the Raman experiment in circular $RL$ polarization reported for $\alpha$-CaCr$_2$O$_4$ in Ref.~\onlinecite{Wulferding_2012}.}
\label{fig:fig3}
\end{figure*}

Raman intensity is highly sensitive to the polarization direction of the incident and the outgoing light in crystals~\cite{JPSJ.23.490,Sugawara_1993,Suzuki_1993,Vernay_2007,PhysRevB.77.174412,PhysRevB.87.174423}. The Raman scattering operator for TLAF contains weight coefficients which vary based on the polarization direction. The dependence on polarization can be expressed as a matrix~\cite{PhysRevB.103.024417}. Based on the choice of the coordinate system established in Fig.~\ref{fig:fig1}, the site locations for the magnetic Cr$^{3+}$ ions of $\alpha$-SrCr$_2$O$_4$ and $\alpha$-CaCr$_2$O$_4$ within the $(b,c)$ plane belongs to the $P_{mmn}$ (D$_{2h}$) symmetry. This creates a polarization coefficient matrix $P_{ij} (\theta,\phi)$ which is given by~\cite{PhysRevB.103.024417}
\begin{equation}
\label{eq:coeff}
\begin{split}
P_{ij}(\theta,\phi)&=\epsilon_{in}(\theta)
\begin{pmatrix}
p_2 & 0 & 0 \\
0 & p_1 & 0 \\
0 & 0 & p_3 
\end{pmatrix}
\epsilon^T_{out}(\phi)\eta^{A_1}_{ij} \\
&+\epsilon_{in}(\theta)
\begin{pmatrix}
0 & 0   & 0    \\
0 & 0   & p_4  \\
0 & p_4 & 0 
\end{pmatrix}
\epsilon^T_{out}(\phi)\eta^{A_2}_{ij}.
\end{split}
\end{equation}
In Fig.~\ref{fig:fig3}(a) we show the scattering geometry that is used to perform the calculation. The incident and outgoing light vectors are expressed as $\epsilon_{in}(\theta)=(0,\cos(\theta),\sin(\theta))$ and $\epsilon_{out}(\phi)=(0,\cos(\phi),\sin(\phi))$.

The total polarized Raman scattering operator $\hat{O}$ is given by
\begin{equation}
\begin{split}
\hat{O}\left(\theta,\phi\right)&=\sum\limits_{i,j}P_{ij}(\theta,\phi)J_{ij}\bs_i\cdot \bs_j \\ &=\hat{O}_2\left(\theta,\phi\right)+\hat{O}_3\left(\theta,\phi\right).
\end{split}
\end{equation}
In the above expression, the polarized bimagnon and trimagnon Raman scattering operators $\hat{O}_2$ and $\hat{O}_3$, respectively, include only the nearest-neighbor and the next-nearest neighbor interactions. The expressions for $\hat{O}_2$ and $\hat{O}_3$ are given by
\begin{equation}
\begin{split}
\hat{O}_2\left(\theta,\phi\right)&=\sum\limits_{i,j}J^{\prime}_{ij}P_{ij}\left(\theta,\phi\right)S\left[\frac{1}{2}\left(-a_ia_j+a_ia^{\dagger}_j+h.c.\right)\right. \\
&+\frac{1}{2}\cos\left(\bq\cdot \bR_{ij}\right)\left(a_ia_j+a_ia^{\dagger}_j+h.c.\right) \\
&\left.-\cos\left(\bq\cdot \bR_{ij}\right)\left(a^{\dagger}_ia_i+a^{\dagger}_ja_j\right)\right],
\end{split}
\end{equation}
\begin{equation}
\begin{split}
\hat{O}_3\left(\theta,\phi\right)&=\sum\limits_{i,j}J^{\prime}_{ij}P_{ij}\left(\theta,\varphi\right)\sqrt{\frac{S}{2}}\sin\left(\bq\cdot \bR_{ij}\right)\times \\ 
&(a_ia^{\dagger}_ja_j+a^{\dagger}_ia^{\dagger}_ja_j-a^{\dagger}_ia_ia_j-a^{\dagger}_ia_ia^{\dagger}_j),
\end{split}
\end{equation}
where the in-plane interaction $J^{\prime}_{ij}$ includes $J_{ch1}$, $J_{ch2}$, $J_{zz1}$, $J_{zz2}$ and $J_{nnn}$. After Fourier transformation, we obtain
\begin{equation}
\begin{split}
\hat{O}_2\left(\theta,\phi\right)&=\sum\limits_{i,j\in u.c.}J^{\prime}_{ij}P_{ij}\left(\theta,\phi\right)S\sum\limits_{\kk}\times \\
&\left[\frac{1}{2}\left(-e^{i\kk\cdot \bR_{ij}}a_{\kk i}a_{-\kk j}+e^{i\kk\cdot \bR_{ij}}a_{\kk i}a^{\dagger}_{\kk j}+h.c.\right)\right. \\
&+\frac{1}{2}\cos\left(\bq\cdot \bR_{ij}\right)\left(e^{i\kk\cdot \bR_{ij}}a_{\kk i}a_{-\kk j}+e^{i\kk\cdot \bR_{ij}}a_{\kk i}a^{\dagger}_{\kk j}+h.c.\right) \\
&\left.-\cos\left(\bq\cdot \bR_{ij}\right)\left(a^{\dagger}_{\kk i}a_{\kk i}+a^{\dagger}_{\kk j}a_{\kk j}\right)\right],
\end{split}
\end{equation}
and
\begin{equation}
\begin{split}
\hat{O}_3\left(\theta,\phi\right)&=\sum\limits_{i,j\in u.c.}J^{\prime}_{ij} P_{ij}\left(\theta,\phi\right)\sqrt{\frac{S}{2N}} \sin\left(\bq\cdot \bR_{ij}\right)\times \\
&\left(\sum\limits_{\kk_2+\kk_3=\kk_1}e^{i\kk_2\cdot\bR_{ij}}a^\dag_{\kk_1j}a_{\kk_2i}a_{\kk_3j}+\sum\limits_{\kk_1+\kk_2=\kk_3}e^{-i\kk_2\cdot\bR_{ij}}a^\dag_{\kk_1j}a^\dag_{\kk_2i}a_{\kk_3j} \right. \\
&\left.-\sum\limits_{\kk_2+\kk_3=\kk_1}e^{-i\kk_2\cdot\bR_{ij}}a^\dag_{\kk_1i}a_{\kk_2j}a_{\kk_3i}-\sum\limits_{\kk_1+\kk_2=\kk_3}e^{i\kk_2\cdot\bR_{ij}}a^\dag_{\kk_1i}a^\dag_{\kk_2 j}a_{\kk_3i}\right).
\end{split}
\end{equation}
Next we apply the Bogoliubov transformation to derive the final expression for the polarized bimagnon and trimagnon Raman scattering operator. They are given by 
\begin{equation}
\label{eq:O2b}
\hat{O}_2\left(\theta,\phi\right)=\sum\limits_{\kk}\sum\limits_{p,q}M^{pq}_{\kk}\left(b_{\kk\;p}b_{-\kk\;q}+b^\dag_{-\kk\;p}b^\dag_{\kk\;q}\right),
\end{equation}
\begin{equation}
\label{eq:O3b}
\hat{O}_3\left(\theta,\phi\right)=\sum\limits_{\kk,\mathbf{l}}\sum\limits_{p,q,t}N^{pqt}_{\kk\mathbf{l}}\left(b_{\kk\;p}b_{-\kk-\mathbf{l}\;q}b_{\mathbf{l}\;t}+b^\dag_{-\kk\;p}b^\dag_{\kk+\mathbf{l}\;q}b^\dag_{-\mathbf{l}\;t}\right),
\end{equation}
where the expressions for the Raman matrix elements $M^{pq}_{\kk}$ and $N^{pqt}_{\kk\mathbf{l}}$ are supplied in Appendix~\ref{appendix b}. Note, in the polarized trimagnon Raman scattering operator $\hat{O}_3\left(\theta,\phi\right)$, the momentum symbol transforms as $\kk_1\rightarrow\kk$, $\kk_2\rightarrow-\kk-\mathbf{l}$ and $\kk_3\rightarrow\mathbf{l}$, where $\kk$ and $\mathbf{l}$ belong to the first BZ. 
%both the intersection angle between the polarized direction and the direction of the $b$ axis while $\theta$ is the intersection angle for incident light and $\phi$ is the intersection angle for outgoing light. 

\begin{figure}[b]
\centering\includegraphics[width=8.6cm]{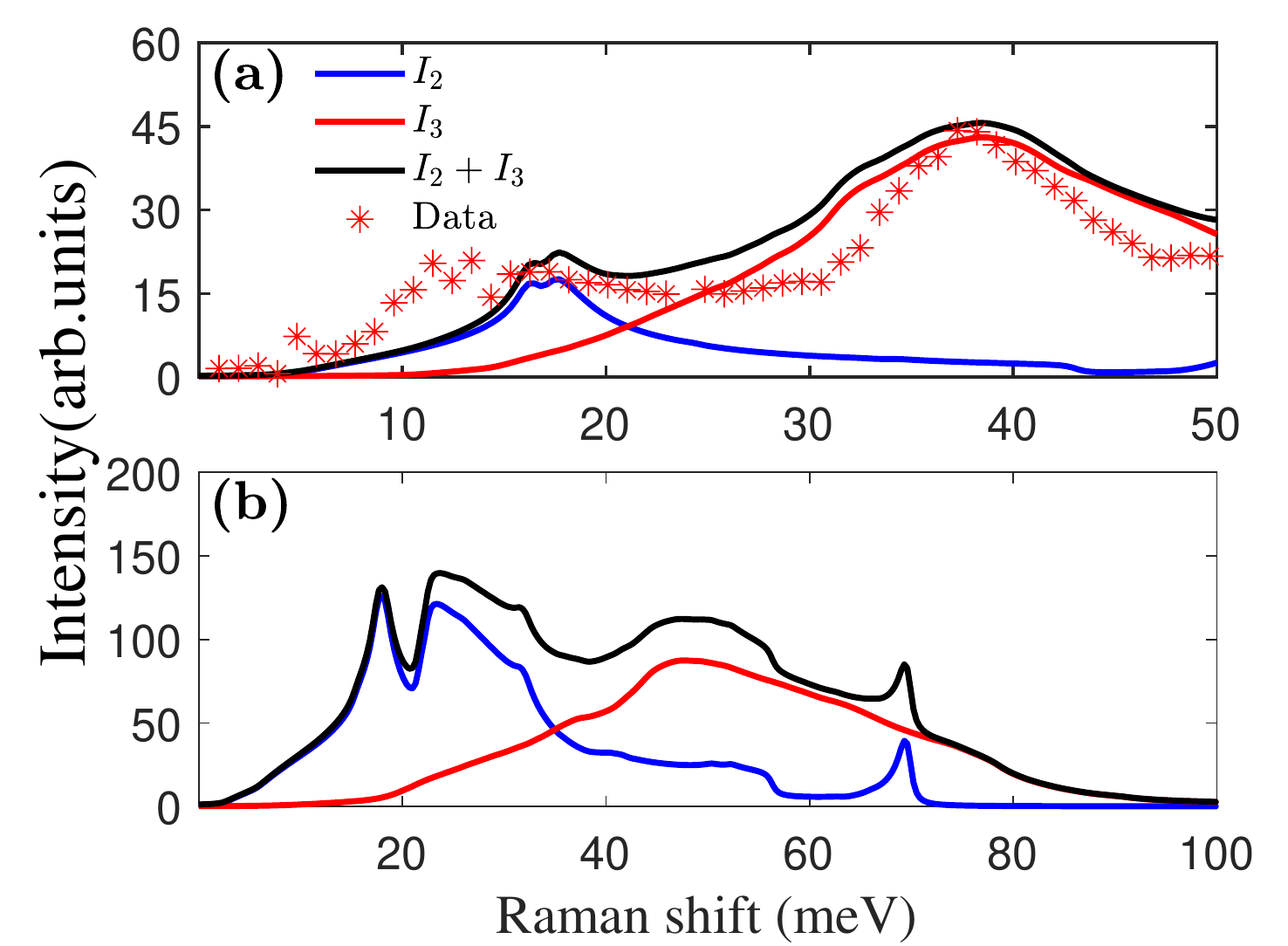}
\caption{Unpolarized Raman spectrum of $\alpha$-SrCr$_2$O$_4$ in (a) and $\alpha$-CaCr$_2$O$_4$ in (b). (a) The blue line is the bimagnon Raman spectrum $I_2$. The red line corresponds to the trimagnon Raman spectrum $I_3$. The black line is the total Raman spectrum $I_2+I_3$. Both $I_2$ and $I_3$ are the results of our fitting procedure. We report a new set of interaction parameters, $\mathcal{P}_6$ in Table.~\ref{table:1}, generated from our model and based on the experimental data reported in Ref.~\onlinecite{PhysRevB.91.144411} (displayed as red asteriks). The calculation utilizes $p_1=p_2=p_3=1$, and $p_4=0.42$. (b) Predicted unpolarized Raman spectrum of $\alpha$-CaCr$_2$O$_4$. The spectrum was calculated based on the experimental  parameter set $\mathcal{P}_5$ reported in Ref.~\onlinecite{PhysRevLett.109.127203}. For $\alpha$-CaCr$_2$O$_4$, $I_2$ and $I_3$ were computed with $p_1=p_2=p_3=p_4=1$.}
\label{fig:fig4}
\end{figure}

\subsection{Polarized Raman spectrum}\label{sec:polarized}
We can use the polarized bimagnon and trimagnon Raman scattering operators introduced in the previous section to construct the bimagnon Green's function $G_2(\omega,\theta,\phi)$ and the trimagnon Green's function $G_3(\omega,\theta,\phi)$. The corresponding definitions are given by
\begin{equation}
\label{eq:g2}
G_2(\omega,\theta,\phi)=-i\int^{+\infty}_{-\infty}dte^{i\omega t}\left<T\hat{O}^\dag_2(t,\theta,\phi)\hat{O}_2(0,\theta,\phi)\right>,
\end{equation}

\begin{equation}
\label{eq:g3}
G_3(\omega,\theta,\phi)=-i\int^{+\infty}_{-\infty}dte^{i\omega t}\left<T\hat{O}^\dag_3(t,\theta,\phi)\hat{O}_3(0,\theta,\phi)\right>.
\end{equation}
%To calculate Green's function for bimagnon and trimagnon excitation, Wick's theorem is applied into polarized bimagnon and trimagnon Raman scattering operators. 
Thus, the polarized bimagnon Raman intensity $I_2(\omega,\theta,\phi)$ and trimagnon Raman intensity $I_3(\omega,\theta,\phi)$ can be written as 
\begin{equation}
\begin{split}
I_2(\omega,\theta,\phi)&=-\frac{1}{\pi}\mathrm{Im}G_2(\omega,\theta,\phi) \\
&=-\frac{1}{\pi}\mathrm{Im}\left[2\sum\limits_{\kk,p,q} \frac{|M^{pq}_{\kk}|^2}{\omega-\omega_{\kk\;p}-\omega_{-\kk\;q}+i0^+}\right],
\end{split}
\end{equation}
and
\begin{equation}
\begin{split}
&I_3(\omega,\theta,\phi)=-\frac{1}{\pi}\mathrm{Im}G_3(\omega,\theta,\phi) \\
&=-\frac{1}{\pi}\mathrm{Im}\left[6\sum\limits_{\kk,\mathbf{l}}\sum\limits_{p,q,t}\frac{|N^{pqt}_{\kk\mathbf{l}}|^2}{\omega-\omega_{\kk\;p}-\omega_{-\kk-\mathbf{l}\;q}-\omega_{\mathbf{l}\;t}+i0^+}\right],
\end{split}
\end{equation}
The polarized Raman intensities depend on the polarization direction of the incident and outgoing lights. In the subsequent calculations, we will fix the incident light polarization angle $\theta$ to be equal to zero and the outgoing light polarization angle $\phi$ will vary according the polarization we choose , see Fig.~\ref{fig:fig3}(a).

The polarized Raman spectrum of the distorted TLAF is shown in Fig.~\ref{fig:fig3}. The vertical axis shows the Raman intensity and the horizontal axis shows the energy re-scaled in units of $\omega/J_{ch1}$. The Raman spectrum for 
$\alpha$-SrCr$_2$O$_4$ was computed using the model described in Sec.~\ref{sec:model} using the parameters stated in Table~\ref{table:1}. The magnetic exchange interactions in $\alpha$-CaCr$_2$O$_4$ are different from those in its Sr counterpart. Whereas the Sr compound has only one $J_{nnn}$, the Ca compound has four. Hence, to keep the calculation tractable for $\alpha$-CaCr$_2$O$_4$ we used an effective $J_{nnn}$ which has only one next-nearest-neighbor exchange interaction as input. This effective value is obtained by taking the average of the four different next-nearest-neighbor exchange energies reported in Ref.~\cite{PhysRevLett.109.127203}.

In the left column of Fig.~\ref{fig:fig3} we show the results of the HH polarization channel. As expected, the bimagnon signals vanish since the Raman scattering operator $\hat{O}_2$ commutes with the Hamiltonian $H_2$. Thus only the trimagnon excitation contributes to the Raman intensity in the HH polarized case. For all the parameter sets that were studied, we observed a pronounced trimagnon continuum vanishing at triple the energy maximum of a single-magnon excitation. We find that as the spatial interaction anisotropy $J_{zz1}-J_{zz2}$ increases the trimagnon continuums in Fig.~\ref{fig:fig3} undergoes a spectral downshift for the HH polarization. 

In the right column of Fig.~\ref{fig:fig3} we show the results of the HV polarization. These plots display the DOS, the bimagnon Raman intensity, and the trimagnon Raman intensity. In HV polarization, we observe both the bimagnon and trimagnon signals. The bimagnon intensity $I_2$ is relatively stronger compared to the trimagnon signal. It occupies an energy region from 0 to approximately 10 for parameter choices of $\mathcal{P}_1$, $\mathcal{P}_2$ and $\mathcal{P}_3$. As an example, consider Fig.~\ref{fig:fig3}(c), where we observe that the $I_2$ signal of the undistorted TLAF shows a peak at 5.82 and an additional peak at 9.36. 

The DOS displays two peaks. A small low energy peak and a strong high energy peak. The DOS signal influences the bimagnon Raman spectrum. At the rotonlike point, the density of magnons is substantially enhanced compared to the other locations in the BZ since $\nabla_\kk E$=0. Since, the velocity vanishes we observe a strong signal in the DOS. From Fig.~\ref{fig:fig3}(c) we notice that this happens at 2.94, which is where the low energy DOS peak is located. This generates the stronger low energy bimagnon Raman intensity. Thus, this signal is a direct consequence of the rotonlike point in the TLAF system. For the undistorted lattice, the 5.82 bimagnon peak arises from both the rotonlike wavevector points M and M$^\prime$. The $I_2$ shows a 9.36 peak due to the one magnon maximum energy which vanishes before 5. 

In an undistorted TLAF he two rotonlike points (M and M$^\prime$) on the Raman spectrum are quivalent. However, as spatial anisotropy is increased  (a fact that naturally occurs in real materials) the Raman spectrum becomes sensitive to this difference. 
Note, in Fig.~\ref{fig:fig3}(g), the anisotropy generates a non-zero DOS contribution at 2.70 from the M$^\prime$ point only. The M point on the other hand has zero contribution (confirmed by calculation). With increasing distortion, the spatial anisotropy changes the spin wave dispersion structure at the M point, thereby causing $\nabla_\kk E \neq 0$. Thus, the M point ceases to behave as a rotonlike point. This is the physical origin of the bimagnon energy peak at 5.34 for $\mathcal{P}_3$. Hence, we see that Raman spectroscopy has the ability to gather insight into the physical nature of the rotonlike point in the TLAF system. 

In Figs.~\ref{fig:fig3}~(g, i, k) the bimagnon signal splits into two peaks when the lattice distortion is enhanced. All the parameter sets from $\mathcal{P}_3$ to $\mathcal{P}_5$, support this peak-splitting effect. The lower energy peak of this two-peak signal originates from the M$^\prime$ rotonlike point. However, it is not obvious what is the physical origin of the higher energy peak within this two-peak signal. Our calculations suggest that the first bimagnon peak for the parameter set $\mathcal{P}_4$ ($\mathcal{P}_5$) in Fig.~\ref{fig:fig3}(i) (Fig.~\ref{fig:fig3}(k)) appears at 1.33 (0.99), which is exactly twice the energy of the local minima shown in Fig.~\ref{fig:fig2}(e) (Fig.~\ref{fig:fig2}(f)).

In addition to distinguishing the different rotonlike points due to the presence of anisotropy, the Raman spectrum has a selective response based on the polarization channels, HH and HV. As noted earlier, the HH channel contribution is exclusively from the trimagnon signal. For the HV channel, the opposite holds true. While there is a weak trimagnon signal, the overwhelming strength comes from the bimagnon excitation. The nearest-neighbor and the next-nearest-neighbor interactions have the same weight along all directions in the trimagnon Raman scattering operator for the HH polarization. However, interactions along the $b$ axis have no contribution on producing trimagnon Raman signal for the HV polarization. Therefore, the trimagnon Raman signal intensity for the HV polarization is observed to be weaker compared to those for HH polarization. Similar to the HH channel, the HV channel also shows a spectral downshift for trimagnon Raman spectrum with increasing distortion.

\subsection{Unpolarized Raman spectrum}\label{sec:unpolarized}
At present there is no consensus on the value of the material parameters that describe the magnetic properties of $\alpha$-SrCr$_2$O$_4$~\cite{PhysRevB.92.214409, PhysRevB.91.144411}, see rows five and six in Table~\ref{table:1}. Motivated by the experimental data of Valentina et al.~\cite{PhysRevB.91.144411}, we compute the unpolarized Raman spectrum of $\alpha$-SrCr$_2$O$_4$. To calculate the unpolarized spectrum we integrate the incident and outgoing light polarization angles $\theta$ and $\phi$ in the Green's function. The resulting expression is given by 
\begin{equation}
\begin{split}
G^u_{\epsilon}(\omega)&=-i\int^{+\infty}_{-\infty}dt\int^{\pi}_{0}d\theta\int^{\pi}_{0}d\phi e^{i\omega t}\left<T\hat{O}^\dag_{\epsilon}(t,\theta,\phi)\hat{O}_{\epsilon}(0,\theta,\phi)\right>, \\
I^u_{\epsilon}(\omega)&=-\frac{1}{\pi}\mathrm{Im}G_{\epsilon}\left(\omega\right).
\end{split}
\end{equation}
where $I_{\epsilon}(\omega)$ represents the unpolarized Raman spectrum with $\epsilon$=2 and 3 referring to the bimagnon and the trimganon Green function channels. The results are presented in  Fig.~\ref{fig:fig4}(a). 

Upon fitting the experimental data of Valentina \emph{et al.} we obtain a new set of parameters that are presented in $\mathcal{P}_6$ in Table.~\ref{table:1}. This new data set reproduces the spin wave dispersion (not shown in the article) of $\alpha$-SrCr$_2$O$_4$, including capturing high energy magnon branches. It also adequately reproduces the experimental unpolarized Raman spectrum of $\alpha$-SrCr$_2$O$_4$. Upon fitting the spectrum displays two prominent peaks. One at 20 meV and the other at 40 meV. The 20 meV signal is primarily composed of the bimagnon channel. But, the 40 meV signal is predominantly trimagnon. Overall the trimagnon intensity is stronger than the bimagnon. This feature of the unpolarized spectrum can allow one to distinguish these two different multi-magnon excitations. Competition between magnetic interaction parameters result in an approximate 120$^{\circ}$ spiral order in $\alpha$-SrCr$_2$O$_4$ [see calculation details in Appendix~\ref{appendix a}]. Therefore, our calculation suggests that the 15 meV Raman shift signal originates from the bimagnon excitation. The signal on the 40 meV Raman shift arises from the trimagnon excitation rather than bimagnon. In Fig.~\ref{fig:fig4}(b), we present our prediction of the unpolarized Raman spectrum of $\alpha$-CaCr$_2$O$_4$. Compared to the Sr-compound, the Ca compound exhibits more significant peak-splitting effect on the bimagnon Raman signal. 

\section{Conclusion}\label{sec:conclusion}
Motivated by the Raman experimental data of TLAF, we investigate the consequences of the non-trivial rotonlike point on the bi- and tri-magnon excitation spectrum of $\alpha$-SrCr$_2$O$_4$ and $\alpha$-CaCr$_2$O$_4$. Utilizing Raman experimental data for $\alpha$-SrCr$_2$O$_4$, we fit our model to propose a new set of magnetic interaction parameters. This new set of parameters is able to consistently reproduce both the inelastic neutron scattering spectrum and the Raman spectrum of $\alpha$-SrCr$_2$O$_4$. Based on our calculations we demonstrate that Raman spectroscopy is sensitive to the behavior of the rotonlike M and M$^{\prime}$ points that have been proposed to exist in the TLAF systems. Additionally, we find that the polarization sensitivity of the incoming beam in Raman spectrum can allow one to distinguish the multimagnon excitation channels (bimagnon versus trimagnon). We observe that the trimagnon Raman signal is the higher energy peak compared to the bimagnon signal. Furthermore, with increasing distortion the peak-splitting effect becomes more prominent in the bimagnon Raman signal.

According to our calculation, the 40 meV Raman signal in the experimental Raman scattering data originates from the trimagnon excitation. The bimagnon excitation could also have some minor contribution to the 40 meV signal. It is worth noting that our proposed magnetic interaction parameters are obtained from fitting with the experimental data of unpolarized Raman spectrum of $\alpha$-SrCr$_2$O$_4$. Nevertheless, the proposed set of new parameters still gives the approximate 120$^{\circ}$ spiral order. This is further validation of our fitting procedure. We also predict the unpolarized Raman spectrum of $\alpha$-CaCr$_2$O$_4$. Compared to the Sr compound, $\alpha$-CaCr$_2$O$_4$ has more intense lattice distortion and stronger in-plane and interlayer exchange interactions. We hope that our theoretical investigation will motivate the TLAF community to further study the connection between the multimagnon excitation and rotonlike points of the TLAF. 

\begin{acknowledgements}
We thank Natalia Drichko for sharing the unpolarized Raman experimental data on $\alpha$-SrCr$_2$O$_4$ and Manila Songvilay for helpful discussions about magnetic interaction parameters for $\alpha$-SrCr$_2$O$_4$. We acknowledge Meiyu He and Chao Shan for useful discussions. This project is supported by NKRDPC-2022YFA1402802, NKRDPC-2018YFA0306001, NSFC-92165204, NSFC-11974432, Shenzhen International Quantum Academy (Grant No. SIQA202102), and Leading Talent Program of Guangdong Special Projects (No. 201626003). T. D. acknowledges the funding support from Sun Yat-Sen University Grant No. OEMT-2019-KF-04 and the hospitality of KITP at UC-Santa Barbara. A part of this research was completed at KITP and was supported in part by the National Science Foundation under Grant No. NSF PHY-1748958. 
\end{acknowledgements}
\begin{appendix}
\section{CLASSICAL GROUND STATE ANALYSIS}\label{appendix a}
Classical ground state energy for $\alpha$-SrCr$_2$O$_4$ is given by
\begin{equation}
E_0\left(\bq\right)=NS^2\sum_{i,j}J_{ij}\cos\left(\bq\cdot\bR_{ij}\right).
\end{equation}
The stabilized spiral order results in
\begin{equation}
\begin{split}
\label{eq:dE0}
\frac{\partial E_0}{\partial Q_b}&=-NS^2\left[J_{ch1}\sin\left(\frac{1}{2}Q_b\right)+J_{ch2}\sin\left(\frac{1}{2}Q_b\right)\right. \\
&\left.+J_{zz1}\sin\left(\frac{1}{4}Q_b\right)+J_{zz2}\sin\left(\frac{1}{4}Q_b\right)+6J_{nnn}\sin\left(\frac{3}{4}Q_b\right)\right]=0.
\end{split}
\end{equation}
By applying the magnetic interaction parameters $\mathcal{P}_6$ in Table.~\ref{table:1}, the solution to Eq.~\eqref{eq:dE0} is $Q_b/2\pi$=1.3031, which is close to the experimental value in Ref.~\cite{PhysRevB.96.024416}. We thus keep using $Q_b/2\pi$=1.3217 for $\mathcal{P}_6$.

\section{POLARIZED RAMAN SCATTERING MATRIX ELEMENTS}\label{appendix b}
The detailed expressions for $M^{pq}_{\kk}$ and $N^{pqt}_{\kk\mathbf{l}}$ in Eq.~\eqref{eq:O2b} and Eq.~\eqref{eq:O3b} are given below
\begin{equation}
\begin{split}
M^{pq}_{\kk}&=\frac{1}{2}J_{ij}P_{ij}\left(\theta,\phi\right)S\left[-e^{i\kk\cdot\bR_{ij}}(u_{\kk\;ip}-v_{\kk\;ip})(u_{-\kk\;jq}-v_{-\kk\;jq})\right. \\ 
&+\cos\left(\bq\cdot\bR_{ij}\right)e^{i\kk\cdot\bR_{ij}}(u_{\kk\;ip}+v_{\kk\;ip})(u_{-\kk\;jq}+v_{-\kk\;jq}) \\
&-\cos\left(\bq\cdot\bR_{ij}\right)(u_{\kk\;ip}v_{-\kk\;iq}+v_{\kk\;ip}u_{-\kk\;iq} \\ &\left.+u_{\kk\;jp}v_{-\kk\;jq}+v_{\kk\;jp}u_{-\kk\;jq})\right],
\end{split}
\end{equation}
\begin{equation}
\begin{split}
N^{pqt}_{\kk\mathbf{l}}&=\frac{1}{6}J_{ij}P_{ij}\left(\theta,\phi\right)\sin\left(\bq\cdot\bR_{ij}\right)\sqrt{\frac{S}{2N}}\times \\
&\left[e^{i\kk\cdot\bR_{ij}}(u_{\kk\;ip}+v_{\kk\;ip})(u_{-\kk-\mathbf{l}\;jq}v_{\mathbf{l}\;jt}+u_{\mathbf{l}\;jq}v_{-\kk-\mathbf{l}\;jt})\right. \\
&-e^{-i\kk\cdot\bR_{ij}}(u_{\kk\;jp}+v_{\kk\;jp})(u_{-\kk-\mathbf{l}\;iq}v_{\mathbf{l}\;it}+u_{\mathbf{l}\;iq}v_{-\kk-\mathbf{l}\;it}) \\
&+e^{i\left(-\kk-\mathbf{l}\right)\cdot\bR_{ij}}(u_{-\kk-\mathbf{l}\;ip}+v_{-\kk-\mathbf{l}\;ip})(u_{\mathbf{l}\;jq}v_{\kk\;jt}+u_{\kk\;jq}v_{\mathbf{l}\;jt}) \\
&-e^{-i\left(-\kk-\mathbf{l}\right)\cdot\bR_{ij}}(u_{-\kk-\mathbf{l}\;jp}+v_{-\kk-\mathbf{l}\;jp})(u_{\mathbf{l}\;iq}v_{\kk\;it}+u_{\kk\;iq}v_{\mathbf{l}\;it}) \\
&+e^{i\mathbf{l}\cdot\bR_{ij}}(u_{\mathbf{l}\;ip}+v_{\mathbf{l}\;ip})(u_{\kk\;jq}v_{-\kk-\mathbf{l}\;jt}+u_{-\kk-\mathbf{l}\;jq}v_{\kk\;jt}) \\
&-\left.e^{-i\mathbf{l}\cdot\bR_{ij}}(u_{\mathbf{l}\;jp}+v_{\mathbf{l}\;jp})(u_{\kk\;iq}v_{-\kk-\mathbf{l}\;it}+u_{-\kk-\mathbf{l}\;iq}v_{\kk\;it})\right].
\end{split}
\end{equation}

\end{appendix}

\bibliography{ref}

\end{document}